\documentclass[twoside]{article}
\usepackage{graphicx,amssymb,mathrsfs,amsmath,color,fancyhdr}
\input vatola.sty \input cyracc.def
\input psfig.sty

\TagsOnRight

\renewcommand{\baselinestretch}{1.06} 
\catcode`@=11 \long\def\@makefntext#1{\noindent #1}
\newskip\tabcentering \tabcentering=1000pt plus 1000pt minus 1000pt
\def\REF#1{\par\hangindent\parindent\indent\llap{#1\enspace}}
\def\MCH#1#2{\setbox0=\hbox{\raise#1\hbox{#2}}\smash{\box0}}

\def\dl{\displaystyle}
\let\@oddfoot\@empty  \let\@evenfoot\@empty

\def\@evenhead{}\def\@oddhead{}
\def\@evenhead{\vbox{\hbox to \textwidth{\footnotesize\rm\hbox to
1.0cm{\thepage\hfill} \hfill\hspace{2mm}\footnotesize{
\emph{Zhang Q F et al}}}}}
\def\@oddhead{\vbox{\hbox to \textwidth{\footnotesize
{\it Error estimate of the two-scale homogenization} \hfill{\ } \hfill\hbox to
1cm{\hfill\thepage}}}}

\def\sec#1{\vspace{2mm}\noindent{{\bf #1}}\vspace{0.5mm}}
\def\th#1{\vspace{1mm}\noindent{\bf #1}\quad } 

\def\leq{\leqslant}

  \def\hml{\end{document}}  \newsymbol\wjzhml 203F \def\no{\noindent}

\begin{document}
\abovedisplayskip=3pt plus 1pt minus 1pt 
\belowdisplayskip=3pt plus 1pt minus 1pt 

\def\le{\leqslant}
\def\ge{\geqslant}
\def\dl{\displaystyle}




\vspace{8true mm}

\renewcommand{\baselinestretch}{1.9}\baselineskip 19pt

\noindent{\LARGE\bf Error estimate of the second-order homogenization for divergence-type nonlinear elliptic equation
with small periodic coef\mbox{}f\mbox{}icients}

\vspace{0.5 true cm}

\noindent{\normalsize\sf ZHANG QiaoFu$^{\dag}$ \& CUI JunZhi
\footnotetext{\baselineskip 10pt
$^\dag$ Corresponding author\\
}}

\vspace{0.2 true cm}
\renewcommand{\baselinestretch}{1.5}\baselineskip 12pt
\noindent{\footnotesize\rm $
 Academy\, of\, Mathematics\, and\, Systems\, Science, \,
Chinese\, Academy\, of \,Sciences, \,Beijing\, 100190,\,China
$ \\
(email: zhangqf@lsec.cc.ac.cn,\,cjz@lsec.cc.ac.cn)\vspace{4mm}}

\baselineskip 12pt \renewcommand{\baselinestretch}{1.18}
\noindent{{\bf Abstract}\small\hspace{2.8mm} 
Second-order two-scale expansions, a unif\mbox{}ied proof for the regularity of the correctors based on the translation invariant
 and a lemma for extracting $O(\varepsilon)$ from the remainder term
are presented for the second order nonlinear elliptic equation with rapidly oscillating coef\mbox{}f\mbox{}icients. If the data are smooth enough,
the error of the zero-order (energy) in $L^\infty$, f\mbox{}irst-order in the H\"older norm, second-order's gradient (flux) in the maximum norm(linear periodic case),
are locally $O(\varepsilon)$.  It can be used in the parabolic equation.
}

\vspace{1mm} \no{\footnotesize{\bf Keywords:\hspace{2mm}}
homogenization,
translation invariant,
De Giorgi-Nash  estimate, error estimate
}

\no{\footnotesize{\bf MSC(2000):\hspace{2mm}}  35B27,\,35J65
 \vspace{2mm}
\baselineskip 15pt
\renewcommand{\baselinestretch}{1.22}
\parindent=10.8pt  
\rm\normalsize\rm

\sec{1\quad Introduction}

Consider the homogenization of the following elliptic problem: find $u_\varepsilon\in H^1_0(\Omega)$,
\begin{equation}\label{eq:fangcheng}
-\frac{\partial}{\partial x_i}\left(a_{ij}(u_\varepsilon,x,\frac x \varepsilon)\frac{\partial u_\varepsilon}{\partial x_j}\right)=f(u_\varepsilon,x,\frac x \varepsilon),\quad \mbox{in\,}\,\Omega\,;
\end{equation}
where $\Omega\subset \mathbb{R}^n$ is  a bounded Lipschitz domain and the summation convention is used.
$A_\varepsilon=(a_{ij})$ is symmetric and positive definite; $a_{ij}(u_\varepsilon,x,y)$ are 1-periodic in y.
In the combined conduction-radiation heat transfer$^{[1]}$, $a_{ij}=k_{ij}(x,\frac x \varepsilon)+4u_\varepsilon^3b_{ij}$.
Assume all of the data are smooth enough.

It is difficult to solve this problem numerically because of the rapidly oscillating coef\mbox{}f\mbox{}icients.
One method is the two-scale homogenization. The error estimate in $L^\infty$ ($O(\varepsilon)$) was presented by
J. L. Lions et al $^{[2]}$ and Lin$^{[3]}$. Oleinik et al$^{[4]}$ proved the $O(\varepsilon^{1/2})$ estimate in $H^1$.
  Su et al$^{[5]}$ investigated the quasi-periodic problems; Zhang and Cui$^{[1]}$  gave a numerical example for the Rosseland equation.
There are also some other famous methods, such as
 Multiscale Finite Element Method(MFEM$^{[6]}$) and Heterogeneous Multiscale Method(HMM$^{[7]}$).

\sec{2\quad Second-order two-scale expansions}

The periodic cell $Y=[0,1]^n$; $W^1_{per}(Y)\subset \{\varphi\in H^1(Y):\int_Y\varphi=0\}$
consists of the functions with the same traces on the opposite faces of Y.
As the same as the linear case$^{[2]}$, we look for a formal asymptotic expansion of the form
\begin{equation}\label{eq:ueps}
u_{\varepsilon}( x)=u_0(x) +
\varepsilon u_1(x,\frac{x}{\varepsilon}) + {\varepsilon}^2
u_2(x,\frac{x}{\varepsilon}) + ...
\end{equation}
where $u_1(\cdot,y),$ $u_2(\cdot, y)$ is Y-periodic in y. Let $y=\frac{x}{\varepsilon}$, then
$\frac{\partial}{\partial x_i} \rightarrow
\frac{\partial}{\partial x_i} + \frac{1}{\varepsilon}
\frac{\partial}{\partial y_i}.$
Assume the coefficient $A_{\varepsilon}$ in \eqref{eq:fangcheng} has the form
$
A_{\varepsilon} =A(u_0,x,\frac{x}{\varepsilon}) + \varepsilon A_1(u_0,u_1,x,\frac{x}{\varepsilon}) + {\varepsilon}^2 A_2 + ...
$
where $A=(a_{ij}(u_0,x,\frac{x}{\varepsilon}))$ is symmetric  positive definite and $A_1,A_2$ are uniform bounded; the righthand side has the form
$
f(u_\varepsilon,x,\frac x \varepsilon)=f(u_0,x,\frac x \varepsilon)+\varepsilon f_1+...
$
with $f_1\in L^\infty$. Substituting \eqref{eq:ueps} into \eqref{eq:fangcheng} and
 equating the power-like terms of $\varepsilon$, we introduce the following auxiliary functions to make the term of order $\varepsilon^{-1}$ equal zero
\begin{equation}\label{eq:n}
\int_Y a_{ij}(u_0,x,y)\frac{\partial N_m}{\partial y_i}\frac{\partial \varphi}{\partial y_j}
=
-\int_Y a_{mj}(u_0,x,y)\frac{\partial \varphi}{\partial y_j},\quad\forall \varphi(y)
\in W^1_{per}(Y), 1\leq m\leq n;
\end{equation}
where $N_m(u_0,x,y)\in W^1_{per}(Y)(u_0,x$ are parameters). Then $u_1=N_l\partial_l u_0$.
The problem for the part of order $\varepsilon^{0}$ admits a unique solution iff there exists $u_0\in  H^1_0(\Omega)$ such that
\begin{equation}\label{eq:u0}
-\frac{\partial}{\partial x_i}
 [a_{ij}^0
\frac{\partial u_0}{\partial x_j}]
 =
\int_Y f(u_0,x,y)\mbox{d} y,\quad a_{ij}^0(u_0,x)
=\int_Y[a_{ij}(u_0,x,y)+a_{il}(u_0,x,y)\frac{\partial N_j}{\partial y_l}]\mbox{d} y.
\end{equation}
This system is well-posed because of the compensated compactness$^{[8]}$.
$\forall \varphi(y)
\in W^1_{per}(Y),$
\begin{eqnarray}\label{eq:m}
&&\int_Y a_{ij}(u_0,x,y)\frac{\partial}{\partial y_i}M_{kl}(u_0,x,y)\frac{\partial}{\partial y_j} \varphi(y)\nonumber\\
&=&
\int_Y \left[a_{kl}+a_{km}\frac{\partial N_l}{\partial y_m}
-\int_Y(a_{kl}+a_{km}\frac{\partial N_l}{\partial y_m})\right]\varphi(y)-
 \int_Ya_{km}N_l\frac{\partial \varphi}{\partial y_m}.
\end{eqnarray}
\begin{eqnarray}\label{eq:q}
\int_Y a_{ij}(u_0,x,y)\frac{\partial Q_{k}}{\partial y_i}
\frac{\partial \varphi}{\partial y_j}
&=&-\int_Y
(a_{il}\frac{\partial N_k}{\partial x_l}+A_{1,ik}+
 A_{1,il}\frac{\partial N_k}{\partial y_l}) \frac{\partial \varphi}{\partial y_i}
\nonumber\\
&&+\int_Y \frac{\partial}{\partial x_i}\left[a_{ik}+a_{il}\frac{\partial N_k}{\partial y_l}
-\int_Y(a_{ik}+a_{il}\frac{\partial N_k}{\partial y_l})
\right]\varphi(y)
.
\end{eqnarray}
\begin{equation}\label{eq:r}
\int_Y a_{ij}(u_0,x,y)\frac{\partial R}{\partial y_i}
\frac{\partial \varphi}{\partial y_j}
=\int_Y[f(u_0,x,y)-\int_Yf(u_0,x,y)]
\varphi(y).
\end{equation}
If $\varphi(y)=1$, the righthand sides of the above equations equal zero. So
 $M_{kl},Q_k,R\in$
 $ W^1_{per}(Y)($$u_0,$
 $x$ are parameters). Let
$
u_2=M_{kl}\partial^2_{kl} u_0
+Q_k\partial_k u_0+R
$ to make the  $O(\varepsilon^{0})$ term equal zero.

 \sec{3\quad Error estimate in $L^\infty$ and $C^\alpha$}

 For the regularity of the following  periodic problem(the abstract Eq. from\eqref{eq:n},\eqref{eq:m}-\eqref{eq:r}),
it can be unified by the translation invariant.
If $Y_z=Y+z, z\in \mathbb{R}^n$, find $M\in W^1_{per}(Y_z)$ such that
\begin{equation}\label{eq:corrector1}
  \int_{Y_z} A(y)\nabla M\cdot \nabla\varphi
  =\int_{Y_z} \vec{B}(y)\cdot \nabla\varphi+d(y)\varphi,\quad\forall \varphi\in W^1_{per}(Y_z),
\end{equation}
where A, $\vec{B}$, d is Y-periodic and $\int_{Y_z}d=0$. Then $M = N$ in $W^1_{per}(Y)$ after a periodic extension  if $N$ is the solution to \eqref{eq:corrector1} in the cell Y.
So the boundary estimate for N is equivalent to the interior estimate for M.
If  $\vec{B}\in L^q$, $d\in L^{q/2}, q>n$, then $N\in C^\alpha$(Th\,8.24$^{[9]}$).

Notice that Lemma 1.6$^{[4]}$ is right for the case $v=1,$ $u\in W^{1,p}$, $p>1$.
If $g(x,y)\in \hat{L}(\Omega\times \mathbb{R}^n),\int_Yg=0$, then by the help of $W^{-1,p}$ and Meyers estimate(Th\,4.2$^{[2]}$) there exists
\begin{equation}\label{eq:oleinik}
  \overrightarrow{\psi}_\varepsilon(x)\in L^{p}(\Omega;\mathbb{R}^n),\quad \mbox{s.\,\,t.}\,\,  g(x,\frac x\varepsilon)
  =\mbox{div}_x \overrightarrow{\psi}_\varepsilon,\quad\|\overrightarrow{\psi}_\varepsilon\|_{L^{p}(\Omega;\mathbb{R}^n)}\leq C\varepsilon.
\end{equation}

Let $Z_\varepsilon\equiv u_\varepsilon- \tilde{u}$$=u_\varepsilon-u_0-\varepsilon u_1-\varepsilon^2u_2$. $[A(u_\varepsilon)-A(\tilde{u})]\nabla
\tilde{u}=Z_\varepsilon A'(\xi)\nabla \tilde{u}\equiv Z_\varepsilon \vec{b}(x,\frac x \varepsilon)$. Then
\begin{equation}\label{eq:z}
   -\frac{\partial}{\partial x_i}\left(a_{ij}(x,\frac x \varepsilon)\frac{\partial Z_\varepsilon}{\partial x_j}
   +b_i(x,\frac x \varepsilon)Z_\varepsilon
   \right)=\varepsilon \frac{\partial \theta_i}{\partial x_i} + \varepsilon \eta,\quad \mbox{in\,}\,\Omega\,,
\end{equation}
because $u_\varepsilon, \tilde{u}$ are known(Th10.7$^{[9]}$).
$Z_\varepsilon=-\varepsilon u_1-\varepsilon^2 u_2
$ is $O(\varepsilon)$ in $L^\infty(\partial\Omega)$ and $O(\varepsilon^{1-\alpha})$ in $C^\alpha(\partial\Omega)$.
If $\|a_{ij}\|_\infty, \|b_i\|_\infty, \|\theta_i\|_q, \|\eta\|_{q/2}\leq C, q>n,$ then $\|Z_\varepsilon\|_\infty, [Z_\varepsilon]_{C^\beta(\Omega')}\leq C\varepsilon$
by the maximum principle and De Giorgi-Nash estimate(Th\,8.16, Th\,8.24$^{[9]}$), $\overline{\Omega'}\subset\subset\Omega$. Since $\|u_1\|_\infty,\|u_2\|_\infty\leq C$,
 $\|u_\varepsilon-u_0\|_\infty\leq C\varepsilon.$ Consequently, for the energy,
 $|\int_\Omega A_\varepsilon\nabla u_\varepsilon\cdot\nabla u_\varepsilon-
  \int_\Omega A^0\nabla u_0\cdot\nabla u_0|\leq C\varepsilon.$
 If $\|\varepsilon u_2\|_{C^\beta}\leq C$, then $\|u_\varepsilon-u_0-\varepsilon u_1\|_{C^\beta(\Omega')}\leq C\varepsilon.$
For the case of nonlinear parabolic equation with mixed boundary conditions, see $^{[10]}$.

If in \eqref{eq:z} $a_{ij}=a_{ij}(\frac x \varepsilon)$, $a_{ij}\in C^\gamma(\overline{Y})$(or piecewise smooth), $b_i=0, \frac{\partial \theta_i}{\partial x_i},
\eta\in L^{n+\delta}(\Omega), \delta>0;$ $\Omega'$ is a open set, $\overline{\Omega'}\subset \Omega.$
Then by the help of Lemma 16$^{[3]}$, $
\sup_{\overline{\Omega'}}|\nabla Z_\varepsilon|\leq
 C\varepsilon.$ If $|\nabla_x u_2|\leq C\varepsilon^{-1},$ then $\sup_{\overline{\Omega'}}|\nabla (u_\varepsilon-u_0-\varepsilon u_1)|\leq
 C\varepsilon.$ It's also true for the tensor case. For the flux, $\sup_{\overline{\Omega'}}|A_\varepsilon\nabla (u_\varepsilon-u_0-\varepsilon u_1)|\leq
 C\varepsilon.$

 Let $w_\varepsilon(x)= [Z_\varepsilon (\varepsilon x)-Z_\varepsilon (0)]/\varepsilon$$^{[3]}$,
 then consider the following equation
 \begin{equation}\label{}
    -\mbox{div}[A(\varepsilon x,x)\nabla w_\varepsilon(x) + \varepsilon w_\varepsilon(x)\vec{b}(\varepsilon x,x)]
    =\varepsilon F(\varepsilon x)
   +Z_\varepsilon(0)\mbox{div}[\vec{b}(\varepsilon x,x)].
 \end{equation}
 where $F(\xi)=\varepsilon \frac{\partial \theta_i}{\partial x_i}(\xi) + \varepsilon \eta(\xi).$
Differentiate both sides formally, by De Giorgi-Nash estimate $[\partial w_\varepsilon]_{C^\gamma(\overline{\Omega'})}\leq C\varepsilon,$ since $|Z_\varepsilon (0)|\leq C\varepsilon$.
 Then locally $[\nabla (u_\varepsilon-u_0-\varepsilon u_1)]_{C^\gamma}\leq C\varepsilon^{1-\gamma}.$

  For the gradient estimate in parabolic case, see Li and Li$^{[11]}$.
 Their work was based on the piecewise smooth coef\mbox{}f\mbox{}icients so it's very important for both theory and practice.

\vspace{3mm}\th{Acknowledgements}
This work is supported by National Natural Science Foundation of China (Grant No. 90916027).
The authors thank Professor Yan NingNing and the referees for their careful reading and helpful comments.

\vskip0.1in \no {\normalsize \bf References}
\vskip0.1in\parskip=0mm \baselineskip 15pt
\renewcommand{\baselinestretch}{1.15}

\footnotesize\parindent=6mm
 \REF{1\ }Zhang Q. F., Cui J. Z.,
Multi-scale analysis method for combined conduction-radiation heat transfer of periodic composites,
 Advances in Heterogeneous Material Mechanics(eds. Fan J. H., Zhang J. Q., Chen H. B., \textit{et al}), Lancaster: DEStech Publications, 2011, 461-464

\footnotesize\parindent=6mm
 \REF{2\ }
Bensoussan A., Lions J. L., Papanicolaou G.,
Asymptotic Analysis for Periodic Structures, Amsterdam: North-Holland, 1978

\footnotesize\parindent=6mm
 \REF{3\ }
Avellaneda M., Lin F. H., Compactness method in the theory of homogenization,
Comm Pure Appl Math, 1987, 40(6): 803-847

\footnotesize\parindent=6mm
 \REF{4\ }
Oleinik O. A., Shamaev A. S., Yosifian G. A.,
Mathematical Problems in Elasticity and Homogenization, Amsterdam: North-Holland, 1992

\footnotesize
\parindent=6mm
 \REF{5\ }Su F., Cui J. Z., Xu Z., \textit{et al},
A second-order and two-scale computation method for the quasi-periodic
structures of composite materials,
Finite Elements in Analysis and Design, 2010,
46(4): 320-327

\footnotesize
\parindent=6mm
 \REF{6\ }Hou T. Y., Wu X. H.,
A Multiscale Finite Element Method for Elliptic Problems in Composite Materials and Porous Media,
 J Comput Phys, 1997, 134(1):169-189

\footnotesize
\parindent=6mm
 \REF{7\ }E W. N., Engquist B., Huang Z. Y.,
Heterogeneous multiscale method: A general methodology for multiscale modeling,
Phys Rev B, 2003, 67(092101): 1-4

\footnotesize\parindent=6mm
 \REF{8\ }Fusco N.,  Moscariello G., On the homogenization of
quasilinear divergence structure operators,
Annali di Matematica Pura ed Applicata, 1986, 146(1): 1-13

\footnotesize\parindent=6mm
 \REF{9\ }Gilbarg D., Trudinger N. S., Elliptic Partial Differential Equations  of Second Order, Berlin: Springer, 2001

\footnotesize\parindent=6mm
 \REF{10\ }
Griepentrog  J. A., Recke L., Local existence, uniqueness and smooth dependence for
     nonsmooth quasilinear parabolic problems, J Evol Equ, 2010, 10(2): 341-375

\footnotesize\parindent=6mm
 \REF{11\ }Li H. G., Li Y. Y., Gradient Estimates for Parabolic Systems from
Composite Material,
arXiv: 1105.1437v1  [math.AP]  7 May 2011
 \hml

\begin{center}
\centerline{\psfig{figure=zkxf33.eps}} \centerline{\footnotesize
Fig. 1.\quad }
\end{center}

{\parbox[c]{60mm}\centerline{\psfig{figure=zkxf33.eps}}
\centerline{\footnotesize Fig. 1.\quad }} {}
{\parbox[c]{60mm}
\begin{center}
\footnotesize Table 1\quad \\\vspace{1.5mm} \doublerulesep 0.4pt
\tabcolsep 19pt
\begin{tabular}{\textwidth}{rcccc}
\hline \hline 表的内容 \hline \hline
\end{tabular}
\end{center}}
}
\newpage
\begin{center}
\footnotesize Table 1\quad \\\vspace{1.5mm} \doublerulesep 0.4pt
\tabcolsep 19pt
\begin{tabular*}{\textwidth}{rcccc}
\hline \hline 表的内容 \hline \hline
\end{tabular*}
\end{center}
\zihao{5}

*******************************************做图*********************
\begin{center}
\centerline{\psfig{figure=zkxf33.eps}} \centerline{\footnotesize
Fig. 1.\quad }
\end{center}

\parbox[c]{60mm}{\centerline{\psfig{figure=zkxf33.eps}}
\centerline{\footnotesize Fig. 1.\quad }}
\parbox[c]{60mm}
{ }
\bf {\boldmath
 \begin{enumerate}
 \item[(1)]
\end{enumerate}
\begin{eqnarray}
 \dot {x} = Ax + Bu,  \quad x(0) = x_0,\nonumber\\
 \dot {x} = Ax + Bu,  \quad x(0) = x_0,
\end{eqnarray}

\def\no{\nonumber}\公式居底
\begin{eqnarray}
\end{eqnarray}

\begin{equation}公式齐缝
\end{equation}
\normalsize
\underline